\documentclass[11pt]{article} 
\usepackage{epsf}
\usepackage{float}
\begin{document}
\baselineskip 16pt plus 2pt minus 2pt
\newcommand{\beq}{\begin{equation}}
\newcommand{\eeq}{\end{equation}}
\newcommand{\beqa}{\begin{eqnarray}}
\newcommand{\eeqa}{\end{eqnarray}}
\newcommand{\dfrac}{\displaystyle \frac}
\renewcommand{\thefootnote}{\#\arabic{footnote}}
\newcommand{\ve}{\varepsilon}
\newcommand{\krig}[1]{\stackrel{\circ}{#1}}
\newcommand{\barr}[1]{\not\mathrel #1}

\begin{titlepage}
 
\hfill {\tiny FZJ-IKP(Th)-2001-03}

\vspace{2cm}

\begin{center}

{\Large  \bf {
Aspects of near threshold neutral pion photoproduction off protons}}

\vspace{1.5cm}
                              
{\large V\'eronique Bernard$^{\ddag}$,
Norbert Kaiser$^{\diamond}$, 
Ulf-G. Mei\ss ner$^{\dag}$}

\vspace{1.2cm}

$^{\ddag}$Universit\'e Louis Pasteur, Laboratoire de Physique
Th\'eorique\\ 3-5, rue de l'Universit\'e, F--67084 Strasbourg, France\\
{\it email: bernard@lpt6.u-strasbg.fr} \\

\vspace{0.4cm}
$^{\diamond}$Technische Universit\"at M\"unchen, Physik Department T39\\ 
James-Franck-Stra{\ss}e, D--85747 Garching, Germany\\
{\it email: nkaiser@physik.tu-muenchen.de}\\

\vspace{0.4cm}
$^{\dag}$Forschungszentrum J\"ulich, Institut f\"ur Kernphysik
(Theorie)\\  D--52425 J\"ulich, Germany\\
{\it email: Ulf-G.Meissner@fz-juelich.de} \\

\end{center}

\vspace{0.7cm}

\begin{abstract}
\noindent We investigate near threshold neutral pion photoproduction 
off protons to fourth
order in heavy baryon chiral perturbation theory in the light of the
new data from MAMI. We show that the unitarity cusp at the secondary $\pi^+n$
threshold is in agreement with expectations from the final state theorem.
We also analyze the fourth order corrections to the P--wave low--energy
theorems and show that potentially large $\Delta$-isobar contributions are 
cancelled by sizeable pion loop effects. This solidifies the parameter free 
third order predictions, which are in good agreement with the data.

\end{abstract}


\vspace{2cm}


\vfill

\end{titlepage}


\section{Introduction}
\label{sec:intro}

Chiral perturbation theory is the tool to systematically investigate the
consequences of the spontaneous and explicit chiral symmetry breaking in 
QCD. S--matrix elements and transition currents of quark operators are
calculated with the help of an effective field theory formulated in terms
of the asymptotically observed fields, the Goldstone bosons and the low--lying
baryons. A systematic perturbative expansion in terms of small external
momenta and meson masses is possible. We call this double expansion from here
on {\em chiral expansion} and denote the small parameters collectively by
$q$. Beyond leading order, coupling constants not fixed by chiral symmetry
appear, the so--called low--energy constants (LECs). These have 
to be determined by a fit to some data or using some model. A large
variety of processes such as pion--nucleon scattering, real and
virtual Compton scattering and so on has already been investigated in
this framework, sharpening our understanding of the chiral structure
of QCD (for reviews,  see e.g.~\cite{bkmrev,misha}).

\medskip
\noindent
Neutral pion photoproduction off protons and deuterons (which gives access to
the elementary neutron amplitudes) is one of the prime processes to test
our understanding of the chiral pion-nucleon dynamics for essentially  two
reasons. First, over the last decade fairly precise differential and total 
cross section data have been obtained at MAMI~\cite{beck,fuchs,berns} and 
SAL~\cite{berg1,berg2,bergd}. A further experiment involving linearly 
polarized photons was performed at MAMI, which not only  improved the 
differential cross sections but also  gave the first determination of the 
photon asymmetry~\cite{schmidt}.
Second, the S-wave amplitude $E_{0+}$ is sensitive to a particular pion
loop effect~\cite{bgkm}. In the threshold region, the fourth
order heavy baryon chiral perturbation theory calculation (HBCHPT) (which
involves the sum $a_1+a_2$ of two low--energy constants) agrees with what is
found in the multipole analysis of the data~\cite{berns,bkmzpc,bkmlet}. In 
addition, the rather counterintuitive prediction for the electric dipole 
amplitude for $\pi^0$ photoproduction off the neutron,  $|E_{0+}^{\pi^0 n}| > 
|E_{0+}^{\pi^0 p}|$ translates into a threshold deuteron amplitude 
$E_d$~\cite{bblmv} that was verified by a SAL experiment within 
20\%~\cite{bergd}. Moreover, in \cite{bkmzpc} it was also  shown that there 
are two {\it P-wave low energy theorems (LETs)} for the $P_{1,2}$
multipoles which show a rapid convergence based on the third order calculation.
While the LET for $P_1$ could be tested and verified from the unpolarized data,
only the recent MAMI measurement of $\vec{\gamma} \,  p \to \pi^0 \, p$ allows
to disentangle the contribution from the $P_2$ and the $P_3$ multipoles (the
latter being largely determined by the LEC $b_P$ at third
order).\footnote{Note that the first but somewhat model--dependent
  comparison between P--wave multipoles and the LET predictions was
  given by Bergstrom~\cite{jack}.} It has been 
frequently  argued that contributions from the delta isobar, that only appear 
at fourth order in the chiral expansion for $P_1$ and $P_2$, will not only 
spoil the rapid convergence of the P--wave LETs but also lead to numerically 
different values. This is witnessed
e.g. in an effective field theory approach including the delta as an active degree
of freedom~\cite{hhk} in which one counts the nucleon--delta mass splitting as
another small parameter. A third order analysis in that framework seems
to indicate indeed large corrections rendering the agreement of the prediction
for $P_1$ at threshold with the value deduced from the differential cross
sections as accidental ~\cite{bhm}.

\medskip
\noindent In this paper, we complete the fourth order (complete one loop)
analysis based on HBCHPT by evaluating the corresponding corrections for the
three P-wave multipoles. We use this framework to analyze the new data from
MAMI, which confirms and sharpens previous findings concerning the electric
dipole amplitude $E_{0+}$ and sheds new light on the convergence issue of the
P--wave LETs.

\medskip

\section{Formal aspects}
\label{sec:form}

In this section, we collect some basic formulas needed for describing the 
reaction $\gamma (k) +p (p_1) \to \pi^0 (q) + p(p_2)$. In the threshold region,
it is  legitimate to consider  $\pi^0$ photoproduction in S-- and 
P--wave approximation, with the corresponding transition current matrix element
given by 
\beq 
{m \over 4 \pi \sqrt s} T\cdot \epsilon = i \vec \sigma \cdot \vec
\epsilon \,\,[E_{0+}(\omega) + \hat k \cdot \hat q\, P_1(\omega)] + i \vec
\sigma \cdot \hat k \, \vec \epsilon \cdot \hat q \, P_2(\omega) + (\hat q
\times  \hat k)\cdot \vec \epsilon\,\, P_3(\omega) ~.
\eeq
Here, $m=938.27$~MeV is the proton mass, $s=(p_1+k)^2=(p_2+q)^2$ the total 
centre-of-mass (cm) energy squared , $\omega=(s-m^2+M_{\pi^0}^2)/2\sqrt{s}$ the
cm energy of the produced neutral pion, and $\epsilon^\mu =(0,\vec \epsilon\,)$
the polarization vector of the real photon in the Coulomb gauge subject to the
transversality condition $\vec \epsilon \cdot \vec k =0$. At threshold, the
$\pi^0$ is produced at rest in the cm frame, $\vec{q} = 0$, so that 
$\omega_{\rm thr} = M_{\pi^0}=134.97$~MeV corresponding to $\sqrt{s}_{\rm thr}
= M_{\pi^0} + m$. The secondary $\pi^+ n$--threshold  opens at  $\omega_c = 
140.11$~MeV where  $\sqrt{s}_{c} =  M_{\pi^+} + m_n$ (with $m_n=939.57$~MeV the
neutron mass). At that point, the strong unitary cusp related to the 
rescattering process $\gamma p \to \pi^+n \to \pi^0 p$ occurs 
in the electric dipole amplitude $E_{0+}(\omega)$. In the vicinity of the cusp
its generic form reads $E_{0+} (\omega) = -a-b \sqrt{1-\omega^2/
\omega_c^2}$ with two constants $a$ and $b$. The amplitudes $P_{1,2,3}(\omega)$
are linear combinations of the more commonly used magnetic dipole ($M_{1\pm}$) 
and electric quadrupole ($E_{1+}$) P--wave pion photoproduction multipoles. The
combinations $P_{1,2,3}(\omega)$ arise most naturally from the decomposition of
the T--matrix in eq.(1). Of importance for the later discussion are also the 
threshold P--wave slopes $\overline P_{1,2}$, defined via
\beq 
\overline P_{1,2} = \lim_{\vec q\to 0}{P_{1,2}(\omega) \over |\vec q\,|}~, 
\eeq 
because for these the LETs mentioned in the introduction have been derived 
in~\cite{bkmzpc}. The differential cross section and the photon asymmetry 
$\Sigma$ can be expressed in terms of the multipoles as
\beqa
{|\vec{k}\,| \over |\vec{q}\,|} {d\sigma \over d\Omega_{\rm cm}} &=&
A + B \, \cos \theta + C\, \cos^2 \theta~, \\
A &=& |E_{0+}|^2 + {1\over 2} |P_2|^2 + {1\over 2} |P_3|^2~, \\
B &=& 2{\rm Re} (E_{0+}P_1^*)~, \\
C &=& |P_{1}|^2 - {1\over 2} |P_2|^2 - {1\over 2} |P_3|^2~,  \\
\Sigma  &=&{|\vec{q}\,| \sin^2 \theta \over 2|\vec{k}\,|}
\biggl( {d\sigma \over d\Omega_{\rm cm}} \biggr)^{-1}~  \Big(|P_3|^2 -  |P_2|^2
\Big)~,  \eeqa
with $\theta$ the cm scattering angle and we have dropped the argument
$\omega$. 

\section{Chiral expansion of the multipoles}
\label{sec:exp}
We wish to calculate the T--matrix element, eq.(1), to order ${\cal O}(q^4)$. 
For the electric dipole amplitude $E_{0+}(\omega)$, this has been already done
in \cite{bkmzpc}. In that paper, also the third order terms for the
P--wave multipoles $P_{1,2,3}(\omega)$ were evaluated. Here, we give these
up-to-and-including fourth order. We make use of the standard heavy baryon 
effective chiral Lagrangian, which has been given to complete one loop accuracy,
i.e. ${\cal O}(q^4)$, in ref.\cite{fmms}. The expressions
for the multipoles split in three parts. First, one has 
the (renormalized) Born terms
which subsume the lowest order couplings ($g_{\pi N}, m$) complemented by the
anomalous magnetic moment ($\kappa_p$) contributions together with all
pion-loop corrections of these parameters at order ${\cal O}(q^3)$ and ${\cal
O}(q^4)$. Secondly, there are the pion loop graphs with at most one insertion
from the dimension two Lagrangian ${\cal L}_{\pi N}^{(2)}$. For these, the
one-nucleon reducible parts which just renormalize the Born terms are taken
out. Thirdly, there are the one-nucleon irreducible counterterm contributions
which lead to simple polynomial amplitudes.   

\medskip
\noindent
Consider first the renormalized Born terms which are expressed in terms of the
physical parameters $g_{\pi N},m,\kappa_p$. The second and third order terms
for $P_{1,2,3}^{\rm Born}(\omega)$  are given in the appendix of
ref.\cite{bkmzpc}. We display here only the novel fourth order contributions,
\beqa
P_1^{\rm{Born}}(\omega) &=& {e g_{\pi N} \,|\vec q\,| \over 320\pi m^4 }\bigg\{
{4M_\pi^4 \over \omega^2 }+(21+19\kappa_p) M_\pi^2+ (20+6\kappa_p)\omega^2
\bigg\} \,, \\
P_2^{\rm {Born}}(\omega) &=& {e g_{\pi N}\, |\vec q\,|\over 320\pi m^4 }\bigg\{
{4M_\pi^4 \over \omega^2 }- (43+15\kappa_p)M_\pi^2 -(26+10\kappa_p) \omega^2
\bigg\} \,, \\
P_3^{\rm{Born}} (\omega) &=& {e g_{\pi N}\, |\vec q\,| \over 160\pi m^4 } \Big\{
(3\kappa_p-2)M_\pi^2 -(13+8\kappa_p)\omega^2\Big\} \,, 
\eeqa
with $|\vec q\,|= \sqrt{\omega^2-M_\pi^2}$. From now on $M_\pi=134.97$~MeV  
denotes the
neutral pion mass, $g_{\pi N}=13.1$ is the strong pion--nucleon coupling 
constant and $\kappa_p = 1.793$ the anomalous magnetic moment of the proton.

\medskip
\noindent
Next, we give the P--wave contributions from the fourth order one loop graphs. 
According to their prefactor, $g_A$ or $g_A^3$, these fall into two classes. 
In these loop diagrams charged as well as neutral pions occur in internal 
lines and we have neglected throughout the small mass difference
$M_{\pi^+}-M_{\pi^0}=4.6$~MeV. In sharp contrast to the S--wave amplitude 
$E_{0+}$ (having a strong cusp effect) this approximation is legitimate for 
P--wave amplitudes since their imaginary parts and consequently their cusp 
effects are extremely tiny corrections. The numerical differences which result 
from taking the charged or neutral pion mass for $M_\pi$ should be regarded  as
an intrinsic inaccuracy of our ${\cal O}(q^4)$ calculation.

\noindent
First, we give the analytical expressions for the ${\cal O}(q^4)$-loops 
proportional to $g_A$:
\begin{eqnarray} P_1^{\rm loop}(\omega)&=& {e g_A\, |\vec q\,| \over m(4\pi
 F_\pi)^3 }  \bigg\{
\Big( 3+{8\over 3} \tilde c_4 \Big)\omega^2\ln{M_\pi \over \lambda}+{4\over 9} 
\tilde c_4(6M_\pi^2-5\omega^2)+{M_\pi^2 \over 2} \arcsin^2{\omega \over M_\pi} 
\nonumber \\ && -{8\tilde c_4 \over 3\omega}
(M_\pi^2 -\omega^2)^{3/2} \arcsin{\omega\over M_\pi} +\omega \sqrt{
M_\pi^2 -\omega^2} \arcsin{\omega\over M_\pi} + {2\omega^3 \over \sqrt{
M_\pi^2 -\omega^2}} \arccos{\omega\over M_\pi} \bigg\} \,, \nonumber \\ &&\\ 
P_2^{\rm loop}(\omega)&=& {eg_A\,|\vec q\,| \over m(4\pi F_\pi)^3}\bigg\{-\Big(
2+ {4\over 3}\tilde c_4\Big)\omega^2 \ln{M_\pi \over \lambda} +\omega^2+{8\over
9}  \tilde c_4(3M_\pi^2+2\omega^2)-2\tilde c_4 M_\pi^2 \arcsin^2 {\omega \over 
M_\pi}\nonumber \\ && +{\pi \over 2} \Big[\omega \sqrt{M_\pi^2 -\omega^2}-
M_\pi^2 \arcsin{\omega\over M_\pi} \Big] -\Big[2\omega +{4\tilde c_4 \over 3
\omega} (\omega^2+2M_\pi^2) \Big]\sqrt{ M_\pi^2 -\omega^2} \arcsin{\omega
\over M_\pi} \bigg\} \,,\nonumber \\ && \\
P_3^{\rm loop}(\omega) &=& {e g_A\, |\vec q\,| \over m(4\pi F_\pi)^3 } \bigg\{
-{\pi \over 2} (1+4\tilde c_4 ) \Big[\omega \sqrt{M_\pi^2 -\omega^2} +M_\pi^2 
\arcsin{\omega\over M_\pi}\Big]\bigg\} \,, 
\eeqa
with $F_\pi= 92.4$~MeV the weak pion decay constant, $\kappa_n=-1.913$ the
neutron magnetic moment,  $g_A = g_{\pi N}F_\pi/m=
1.29$ and $\tilde c_4 =m c_4$. The low--energy constant $c_4$ has been
determined from pion--nucleon scattering inside the Mandelstam triangle as 
$ c_4 = 3.4$~GeV$^{-1}$ \cite{paulb}. $\lambda$ is the scale of  
dimensional regularization which will be set equal to $\lambda = m$.
As a check these loop contributions fulfill the condition $P_{1,2,3}^{\rm
loop}(0)= 0$ which confirms that all one-nucleon reducible pieces are
indeed taken out. 
The analytical continuation above threshold $\omega > M_\pi$ is obtained by the
following substitutions:
\beq \sqrt{M_\pi^2-\omega^2} \to -i \, \sqrt{\omega^2-M_\pi^2} \,, \qquad
\arcsin{\omega \over M_\pi} \to {\pi \over 2}+ i \, \ln{\omega
+\sqrt{\omega^2-M_\pi^2} \over M_\pi} \,. \eeq
\noindent Similar analytical expressions are found for the other 
class of ${\cal O}(q^4)$-loops proportional to $g_A^3$,
\begin{eqnarray} 
P_1^{\rm loop}(\omega)&=& {e g_A^3\,|\vec q\,|\over m(4\pi F_\pi)^3 }\bigg\{
{\omega^2\over 3}(7+4\kappa_p+2\kappa_n) \ln{M_\pi \over \lambda} 
-{(M_\pi^2 -\omega^2)^{3/2} \over 3 \omega} (7+4\kappa_p+2\kappa_n)\arcsin
{\omega\over M_\pi}\nonumber \\ && + {\pi M_\pi^2\over 6\omega^3}\bigg[4M_\pi^3
-6\omega^2M_\pi -{3\omega^4\over M_\pi} +3\omega (2\omega^2 -M_\pi^2) \arcsin
{\omega\over M_\pi}+( 7\omega^2 -4M_\pi^2)\sqrt{M_\pi^2 -\omega^2} \bigg] 
\nonumber \\ && +{M_\pi^2\over 6} (17+8\kappa_p+4\kappa_n)
-{\omega^2\over 9} (19+10\kappa_p+5\kappa_n)+{M_\pi^2\over 2\omega^2}(\omega^2
-M_\pi^2) \arcsin^2{\omega\over M_\pi}\bigg\} \,, \\
P_2^{\rm loop}(\omega)&=& {e g_A^3\, |\vec q\,| \over m(4\pi F_\pi)^3 }\bigg\{-
{2 \over 3} (5+2\kappa_p+\kappa_n)\omega^2 \ln{M_\pi \over \lambda}+{\omega^2
\over
9} (19+10\kappa_p+5\kappa_n) \nonumber \\ && +{\sqrt{M_\pi^2-\omega^2}\over 3 
\omega}\Big[M_\pi^2 (7+4\kappa_p+2\kappa_n) -2\omega^2 (5+2\kappa_p+\kappa_n)
\Big] \arcsin {\omega\over M_\pi}-{M_\pi^4\over 2\omega^2} \arcsin^2{\omega
\over M_\pi} \nonumber \\ && + {\pi M_\pi \over 6 \omega^3}\bigg[4M_\pi^4 
-6\omega^2M_\pi^2 +3\omega^4 -4 M_\pi (M_\pi^2-\omega^2)^{3/2}\bigg]-{M_\pi^2
\over 6} (11+8\kappa_p+4\kappa_n) \bigg\}\,, \\
P_3^{\rm loop}(\omega) &=& {e g_A^3\, |\vec q\,|\over m(4\pi F_\pi)^3 } \bigg\{
{\pi \over 3\omega } (\kappa_n-3\kappa_p-3) \Big[M_\pi^3 -(M_\pi^2 -\omega^2)^{
3/2}\Big]\bigg\} \,, 
\eeqa
which also fulfill the nontrivial condition $P_{1,2,3}^{\rm loop}(0)=0$.

\medskip\noindent
Finally, we are left with the polynomial counterterm contributions,
\beqa
P_1^{\rm ct}(\omega) &=& {e g_A \,|\vec q\,| \,\omega^2 \over m(4\pi F_\pi)^3 } \,
\xi_1(\lambda) \,,\\
 P_2^{\rm ct}(\omega) &=& {e g_A \,|\vec q\,| \,\omega^2 \over m(4\pi
F_\pi)^3} \, \xi_2(\lambda) \,, \\
P_3^{\rm ct}(\omega) &=& e \, |\vec q\,|\, b_P \Big\{ 
\omega -{M_\pi^2 \over 2m }\Big\} \,.
\eeqa
The introduced new parameters (LECs) $\xi_{1,2}(\lambda)$ are dimensionless and
they balance of course the scale dependence appearing in the fourth order loop
contribution via the chiral logarithm $\ln(M_\pi/\lambda)$. The form of  
$P_3^{\rm ct}(\omega)$ follows from the relativistic operators $O_8$ and $O_9$
constructed  in ref.\cite{fmms}. The LEC $b_P$ already appeared at third order,
the only new feature here is a kinematical correction $|\vec k\,| = \omega
-M_\pi^2/2m + \dots$, which at threshold amounts to a 7\% reduction.  

\medskip\noindent
Furthermore, we give the resonance contributions to the low energy constants
$\xi_{1,2}$ and $b_p$. As mentioned in ref.\cite{bkmzpc} there is a small
contribution to $b_P$ from t-channel vector meson exchange ($\rho^0(770)$ and
$\omega(782)$) and consequently also an analogous vector meson exchange 
contribution to $\xi_{1,2}$,
\beq\label{PletV} 
b_P^{(V)} = {5\over (4\pi F_\pi)^3} \,, \qquad \xi_1^{(V)}= -2 \xi_2^{(V)}
= -{8 \over g_A}\,. 
\eeq
Here, we have used various simplifying relations (see ref.\cite{bkmzpc}) for 
the vector meson coupling constants together with the KSFR relation for the
vector meson masses $M_\rho \simeq M_\omega$.    
 The dominant contribution to $b_P$ and 
$\xi_{1,2}$ come from the low-lying $\Delta(1232)$ resonance. 
In refs.\cite{bkmzpc,bkmlet} we used a relativistic tree level approach in
which the delta contribution is parametrized in terms of four  couplings $g_1$,
$g_2$, $Y$ and 
$Z$. The latter two are so-called off-shell parameters emerging in a
relativistic description of the spin-3/2 fields. In a corresponding
effective Lagrangian, these would be represented by some higher order
contact interactions. In fact, most of the delta resonance physics can be 
represented by the static isobar approach, which can be thought of as the
leading term in a systematic effective field theory expansion like the one
given in \cite{hhk}. In the study of pion--nucleon scattering~\cite{FM} it was
already demonstrated that the dominant isobar contributions come indeed from 
the lowest order Born graphs. Therefore, we use here the static
non--relativistic isobar--model where  one gets the following expressions,
\beq \label{PletD}
b_P^{(\Delta)} = {\kappa^*\, g_A  \over 6 \sqrt{2} \pi m F_\pi}\, {\Delta \over
\Delta^ 2-M_\pi^2} \,, \eeq
\beq \xi_1^{(\Delta)} =-\xi_2^{(\Delta)} =    {\kappa^* \over 3\sqrt{2} }\,\,
 {(4\pi F_\pi)^2 \over \Delta^2 -M_\pi^2 } \,, \eeq
with $\Delta = 293$~MeV the delta-nucleon mass--splitting and $\kappa^*$ the
$N\Delta$ transition magnetic moment.   It is important for the numerical 
evaluation  to keep the $M_\pi^2$-term in the denominator, this is also
justified in the small scale expansion, where one counts $\Delta$ as a small
parameter like the pion mass. 

\medskip\noindent
For the later discussion of the P--wave LETs, we now give the various
contributions to the slopes $\overline{P}_{1,2}$. We start with the
renormalized Born terms expressed by  the physical pion--nucleon coupling
constant $g_{\pi N}$, the renormalized anomalous magnetic moment $\kappa_p$
and the proton mass $m$. Furthermore, we introduce the small parameter $\mu =
M_{\pi^0}/m \simeq 0.144$ and have 
\beqa \label{LETBorn}
\overline P_1({\rm Born}) &=&{e g_{\pi N} \over 8 \pi m^2} \bigg[ 1 +
\kappa_p - {\mu \over 2} (2+\kappa_p) +{\mu^2 \over 8} (9 + 5 \kappa_p) \bigg]
~, \\ 
\overline P_2({\rm Born}) &=& {e g_{\pi N} \over 8 \pi m^2} \bigg[ -1 -
\kappa_p + {\mu \over 2} (3+\kappa_p) -{\mu^2 \over 8} (13 + 5
\kappa_p) \bigg]~.
\eeqa
Here, novel terms of order $\mu^2$ appear. The result for the chiral loops at 
order ${\cal O}(q^3)$ can be taken from \cite{bkmzpc},
\beqa \label{LETloop3}
\overline P_1(q^3-{\rm loop}) &=&{e g^3_{\pi N}\,\mu\over 384 \pi^2 m^2}
(10-3\pi) ~, \\ \overline P_2(q^3{\rm -loop})&=&-{e g^3
_{\pi N} \mu \over 192 \pi^2 m^2}~. 
\eeqa
These contributions are known to be quite small.
{}From the formulae for $P_{1,2}^{\rm loop}(\omega)$ and $P_{1,2}^{\rm ct}
(\omega)$ given above, one can readily deduce the terms due to the chiral loops
and counterterms at order ${\cal O}(q^4)$, 
\beqa \label{LETloop41}
\overline P_1(q^4{\rm -loop,ct})&=&{e g_A M_\pi^2 \over m (4\pi F_\pi)^3}
\bigg\{ \bigg[ {8\over 3} \tilde c_4 +3 +{g_A^2\over 3} (7+4\kappa_p+2\kappa_n)
\bigg] \ln{M_\pi \over \lambda}  -g_A^2{5\pi\over 6} \nonumber \\ 
&&+(1+2g_A^2){\pi^2\over 8} +{4\over 9} \tilde c_4 +2 +{g_A^2 \over 18}
(13+4\kappa_p+2\kappa_n) +\xi_1(\lambda)\bigg\}~.\\ 
\overline P_2(q^4{\rm -loop,ct})&=&{e g_A M_\pi^2
\over m (4\pi  F_\pi)^3} \bigg\{\bigg[ -{4\over 3} \tilde c_4-2-{2g_A^2\over 3}
(5+2 \kappa_p+ \kappa_n) \bigg] \ln{M_\pi \over \lambda} + g_A^2 {\pi\over 6} 
\nonumber\\&&-(4\tilde c_4+2+g_A^2){\pi^2\over8} +{40\over 9} \tilde c_4 +1 
+{g_A^2 \over 18}(5-4\kappa_p-2 \kappa_n)+\xi_2(\lambda) \bigg\}~. 
 \label{LETloop42}
\eeqa
The resulting numerical values will be given later.

\section{Results and discussion}
\label{sec:res}

The new MAMI differential cross section data span the energy range
from $E_\gamma = 145.1\,$MeV to 165.6~MeV in steps of about 1.1~MeV.
In addition, the photon asymmetry $\Sigma$ has been evaluated for energies
from 145 to 166~MeV, all these data for $\Sigma$ have been binned to one 
average energy of $E_\gamma = 159.5\,$MeV. In total, we have 171 differential
cross section data, 19 total cross section points and 7 data points 
for the photon asymmetry $\Sigma$. 

\noindent
First, is it instructive to compare the new data with the previously 
obtained ones of Fuchs et al.~\cite{fuchs}. For doing that,
we  compare fits using the fourth order expressions for the S--wave amplitude
$E_{0+}$  and the third order ones for the P--wave amplitudes (as it was done
in \cite{bkmzpc,bkmlet}). The resulting LECs and
$\chi^2$/dof are collected in table~\ref{tab:3}.
\renewcommand{\arraystretch}{1.2}
\begin{table}[hbt]
\begin{center}
\begin{tabular}{|l|c|c|}
    \hline
                       &   Schmidt et al. & Fuchs et al. \\ 
    \hline
$a_1$ [GeV$^{-4}$]       &  10.585          &  3.464       \\
$a_2$ [GeV$^{-4}$]       &  $-$4.542        &  3.136       \\
Corr($a_1,a_2$)          &  $-$0.998        & $-$0.999     \\
$a_1+a_2$ [GeV$^{-4}$]   &   6.04           &  6.60        \\
$b_P$ [GeV$^{-3}$]       &  14.84           & 13.00        \\
No. of data              &   190            &  180         \\
$\chi^2$/dof             &   3.19           &  2.20        \\
\hline\hline
\end{tabular}
\centerline{\parbox{11cm}
{\caption{Values of the LECs resulting from a three parameter fit to the 
         cross section data of ref.\protect\cite{fuchs} and  \protect\cite{schmidt}.
         Corr. denotes correlation between the two S--wave LECs.
         \label{tab:3}}}}
\end{center}
\end{table}   
\noindent
In both cases the two S--wave LECs are completely anticorrelated, i.e.
only the sum $a_1 + a_2$ is of relevance. It agrees within 10\% for
the two fits, showing that the S--wave multipole $E_{0+}$ is internally 
consistent.
The P--wave LEC $b_P$ is somewhat increased, but now consistent with the
value obtained from fitting the SAL data~\cite{berg1,berg2}, 
$b_P^{\rm SAL} \simeq 15\,$GeV$^{-3}$~\cite{bkmlet}.
This can simply be traced back to the fact that the new MAMI total cross
sections are  larger than the old ones above the secondary
$\pi^+n$--threshold. It is gratifying that this so far puzzling experimental
discrepancy is now resolved. 

\medskip\noindent Next, we wish to investigate the strength of the S--wave
cusp. For that, we use the realistic two--parameter model developed in
ref.\cite{bkmzpc} (which is similar to the so--called unitary fit of
ref.\cite{berns}), where $E_{0+}$ is given by
\beq
E_{0+} (\omega ) = -a - b \, \sqrt{1 - {\omega^2 \over \omega^2_c}}~.
\eeq
Assuming isospin invariance for $\pi$N--rescattering, the strength of the 
cusp given by the parameter $b=\sqrt 2 \, a^- \, M_\pi \, E_{0+}^{\pi^+n}$ 
can be inferred from the well measured pion-nucleon scattering length 
$a(\pi^- p \to \pi^0 n)$ and the precise CHPT prediction for the electric
dipole amplitude $E_{0+} (\gamma p \to \pi^+ n)$ at threshold (which agrees
with the data). This gives $b = (3.67 \pm 0.14)\cdot 
10^{-3}/M_{\pi^+}$~\cite{berns}. Fitting the older
MAMI data, the resulting value for $b$ came out sizeably smaller, $b\simeq 2.8
\cdot 10^{-3}/M_{\pi^+}$~\cite{bkmlet}. This prompted some speculations that
the strength of the unitary cusp is very sensitive to isospin violation. If,
however, we use this same model together with the third order predictions
for the P--waves and apply it to the new MAMI data, we get
\beqa\label{abf}
a &= & 0.54 \cdot 10^{-3}/M_{\pi^+}~, \nonumber \\
b &= & 3.63 \cdot 10^{-3}/M_{\pi^+}~, \nonumber \\
b_p &=& 14.43 \,\, {\rm GeV}^{-3}~, 
\eeqa
with a $\chi^2$/dof of 3.21, which is of the same quality as the one
of the three parameter HBCHPT fit discussed before. The value for $b$
in eq.(\ref{abf}) is in perfect agreement with the prediction obtained from
the final state theorem and assuming isospin invariance for 
$\pi$N--rescattering. That sheds some
doubt on the speculation that a precise measurement of the unitary cusp
would be a good tool to investigate isospin violation. The resulting
real part of $E_{0+}$ is given by the dash--dotted line in fig.~\ref{fig:E0}.
\begin{figure}[t]
\centerline{
\epsfysize=2.5in
\epsffile{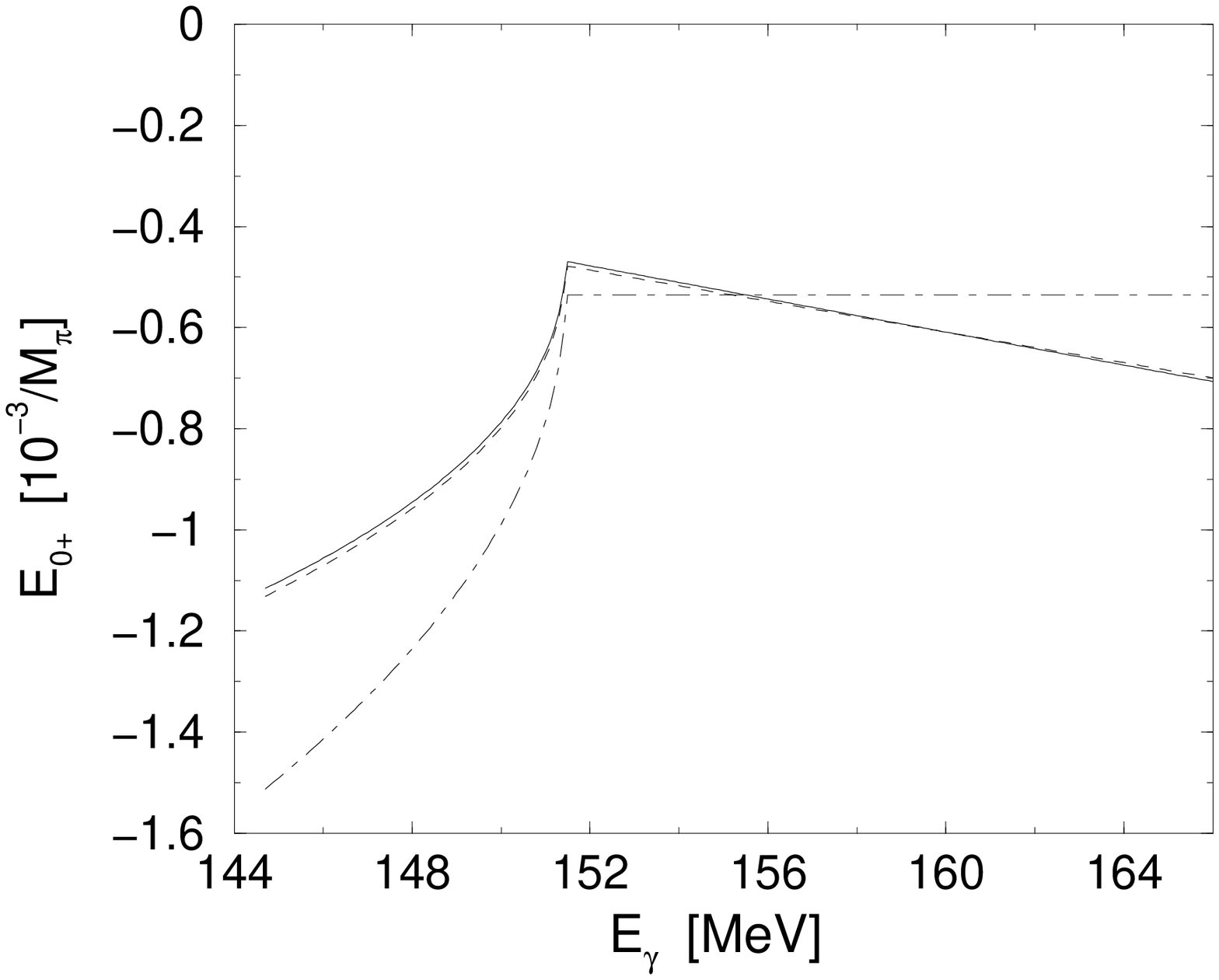}
\hspace{0.5cm}
\epsfysize=2.5in
\epsffile{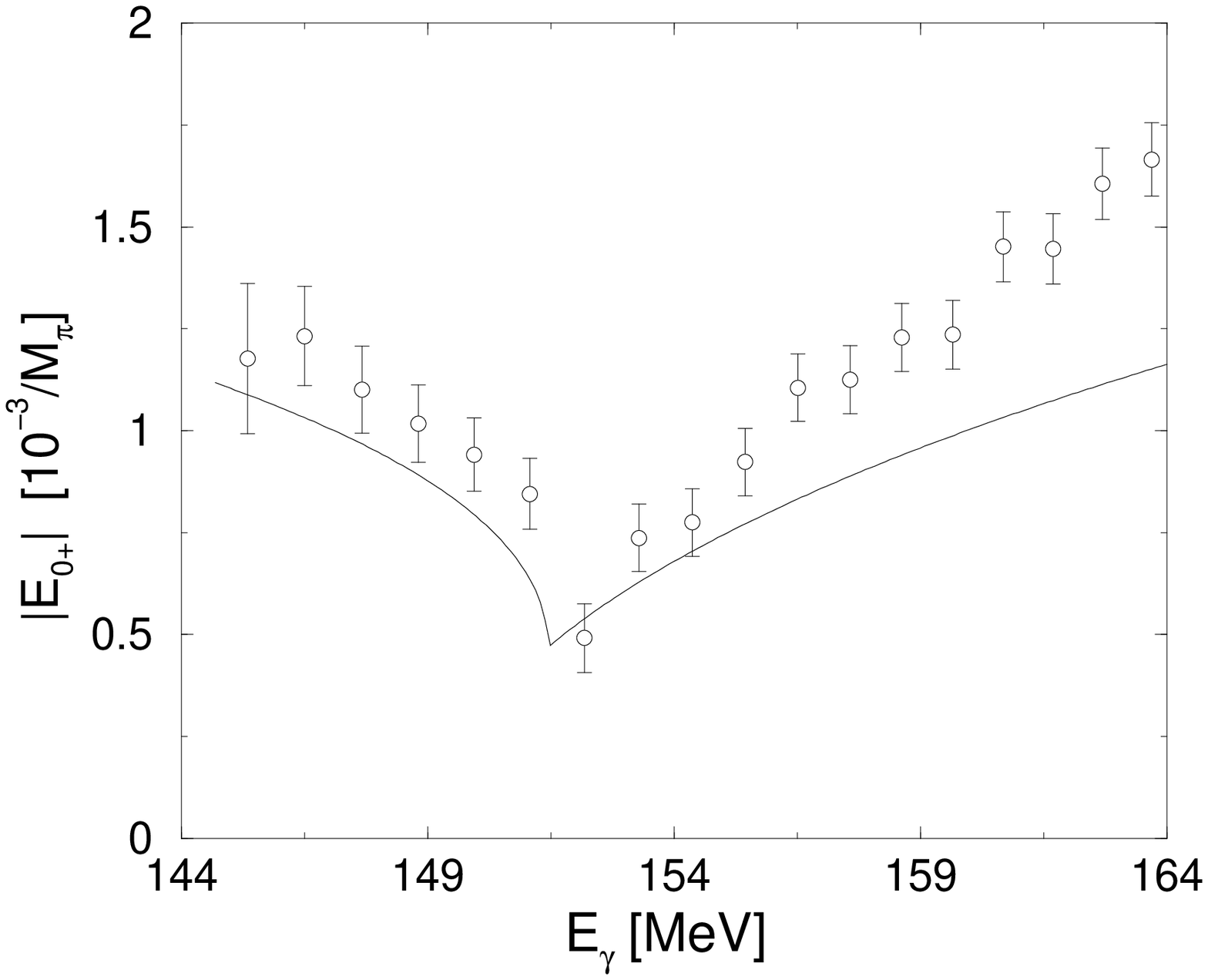}
}
\vskip 0.1cm
\centerline{\parbox{16cm}{\caption{
The electric dipole amplitude in the threshold region. Left panel: Real part.
The various lines are explained in the text. Right panel: The modulus of
$E_{0+}$ for the full fit (see text) in comparison to the 
SAL data~\protect\cite{berg1,berg2}.
\label{fig:E0} 
}}}
\end{figure}   
\noindent In this simple three parameter approach, one can also predict the
photon asymmetry $\Sigma$, since $P_3$ is governed by the LEC $b_P$ and $P_2$
is given by the LET. Of course, in the full fit involving also the fourth order
contribution to the P--waves, one has additional terms from loops and
counterterms, nevertheless, in this simplified ansatz one can already estimate
the corrections to be expected from these additional terms. Using the LECs
collected in table~\ref{tab:3}, 
we obtain the dash-dotted in line fig.~\ref{fig:a}, which agrees 
quite well with the data from the MAMI analysis. Therefore, we
conclude that the corrections to the $P_2$ multipole should be small. Also,
we had already noted before that $P_3$ is only modified on the percent level
by the fourth order corrections. Thus, to keep the fine balance between
$|P_2|^2$ and $|P_3|^2$, which governs the size of $\Sigma$, only modest
corrections to $P_2$ should be expected.
\begin{figure}[H]
\centerline{
\epsfysize=2.5in
\epsffile{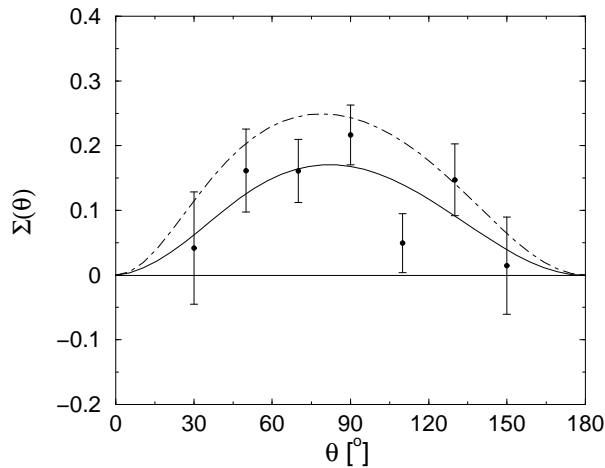}
}
\vskip 0.1cm
\centerline{\parbox{11cm}{\caption{
The photon asymmetry at $E_\gamma = 159.5$~MeV. The 
various lines are explained in the text. 
The data are from \protect\cite{schmidt}.
\label{fig:a}
}}}
\end{figure}

\renewcommand{\arraystretch}{1.2}
\begin{table}[hbt]
\begin{center}
\begin{tabular}{|l|c|c|}
    \hline
                            &   Set~I          &   Set~II     \\ 
    \hline
$a_1$ [GeV$^{-4}$]          &  7.734           &  8.588       \\
$a_2$ [GeV$^{-4}$]          &  $-$1.506        &  $-$2.288   \\
Corr($a_1,a_2$)             &  $-$0.998        & $-$0.998     \\
$a_1+a_2$ [GeV$^{-4}$]      &   6.23           &  6.30        \\
$\chi^2$/dof                &   1.36           &  1.35        \\
$E_{0+} (\omega_{\rm thr})$ [$10^{-3}/M_{\pi^+}$] 
                            &  $-$1.13         & $-$1.12      \\
$E_{0+} (\omega_c)$         [$10^{-3}/M_{\pi^+}$]       
                            &  $-$0.53         & $-$0.52      \\
\hline\hline
\end{tabular}
\centerline{\parbox{11cm}{
\caption{Values of the S--wave LECs and $E_{0+}$ at the two thresholds
         resulting from the five parameter
         fits of the data of ref.\protect\cite{schmidt}.
         Corr($a_1,a_2$) denotes correlation between the two S--wave LECs.
         \label{tab:S}}}}
\end{center}
\end{table}   
\medskip
\noindent We now discuss the full fits including the fourth order
corrections to the P--wave multipoles $P_{1,2,3}(\omega)$. We have performed 
two types of fits.
In set~I, we only fit to the differential and total cross section
data excluding the photon asymmetry $\Sigma$. For set~II, we include the photon
asymmetry data
in the fits. Let us first discuss the electric dipole amplitude.
It should come out (largely) independent of the fitting procedure since
$\Sigma $ is only indirectly sensitive to the S--wave. The
resulting LECs are collected in table~\ref{tab:S}. As expected,
one finds a very similar result for $a_1+a_2$ in agreement with the
previous determinations. The resulting $E_{0+} (\omega)$ comes out
indeed independent of the fitting procedure as shown by the solid
and dashed lines in fig.~\ref{fig:E0}. It is in good agreement with
determinations based on the older MAMI and the SAL data, which
lead to $E_{0+} (\omega_{\rm thr}) = -(1.3 \pm 0.2) \cdot
10^{-3}/M_{\pi^+}$ and also with the result 
obtained from the new MAMI data, $E_{0+} (\omega_{\rm thr}) = 
-(1.33 \pm 0.08 \pm 0.03) \cdot 10^{-3}/M_{\pi^+}$.
Therefore, even though the convergence of the
chiral expansion in this multipole is slow, the fourth order
calculation is able to describe it in the threshold region with
one parameter (the sum of LECs $a_1+a_2$). This small value for
$E_{0+}$ at threshold clearly establishes the large pion loop effect
first pointed out in \cite{bgkm}. For completeness, we also show in
fig.~\ref{fig:E0} the modulus of the electric dipole amplitude, 
$|E_{0+}| = ([{\rm Re}~E_{0+}]^2 + [{\rm Im}~E_{0+}]^2)^{1/2}$, in
comparison to the data from SAL~\cite{berg1,berg2}, which nicely
shows the unitary cusp.

\medskip
\noindent We now turn to the P-waves. Here, we encounter the following
problem. While the best fit of type~I gives a good $\chi^2$/dof, see
table~\ref{tab:S}, there is an almost perfect correlation between
$b_P$, $\xi_1$ and $\xi_2$ (which is expected since the differential cross
sections are only sensitive to $|P_2|^2 +|P_3|^2$) and the resulting
values for $b_P$ or $\xi_2$ come out either too large (based on expectations
from resonance exchange, to be discussed below) or with too large
uncertainty. We have also performed fits with fixing $b_P$ at
the previously determined value of 14.8~GeV$^{-3}$, which gives almost the
same $\chi^2$/dof but a wastly different value for $\xi_2$. Furthermore,
including the leading effects of D--waves in the low energy region does 
not change this. On the other hand, in all cases the LEC $\xi_1$ comes
out in a narrow range, which is in  agreement with the 
estimates based on resonance exchange to be discussed next.
If one includes the photon asymmetry data, the cross
sections and the photon asymmetry are well described, see the solid
line in fig.~\ref{fig:a}, but the resulting value for $b_P$ is too
large whereas $\xi_1$ and $\xi_2$ come out of the size expected from
resonance saturation. For completeness, we show the energy dependence
of the real parts of the P--wave multipoles for the fit 
including the asymmetry data in fig.~\ref{fig:P}.

\medskip

\begin{figure}[htb]
\centerline{
\epsfysize=3in
\epsffile{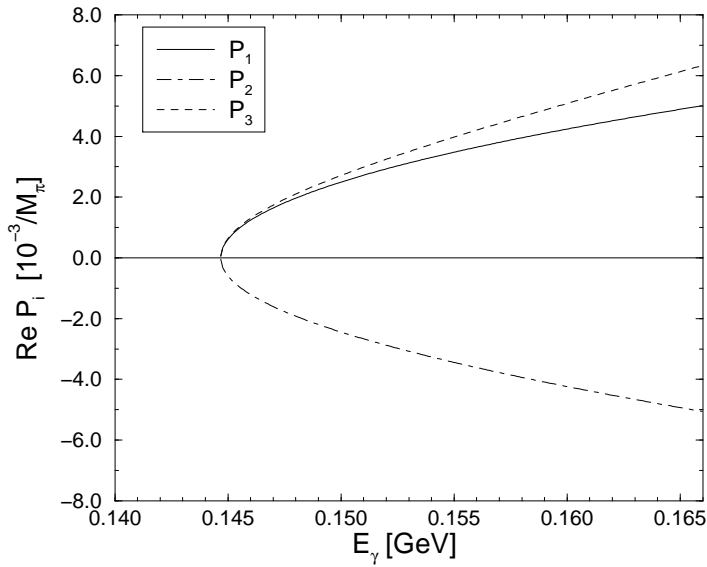}
}
\vskip 0.1cm
\centerline{\parbox{13cm}{\caption{
Real part of the P--waves $P_i$ ($i=1,2,3$) versus photon energy.
\label{fig:P}
}}}
\end{figure}

\medskip\noindent
Therefore, to get a more reliable estimate of the corrections to the P--wave
LETs, we employ the resonance saturation hypothesis. First, taking the 
parameters used here, the LET predictions read (which are nothing but the
sum of the third order renormalized Born and loop terms):
\beq\label{LETt}
\begin{tabular}{lrr}
$\overline P_1^{\rm LET}$ =& $0.469~{\rm GeV}^{-2}$~, 
& $\quad [0.445, 0.492]~{\rm GeV}^{-2}$~, \\
$\overline P_2^{\rm LET}$ =& $-0.498~{\rm GeV}^{-2}$~,
 & $\quad [-0.472, -0.523]~{\rm GeV}^{-2}$~,
\end{tabular}
\eeq
where the numbers in the square brackets refer to a 5\% 
theoretical uncertainty. The
results based on the new MAMI data are~\cite{schmidt}
\beqa\label{LETe}
\begin{tabular}{lr}
$\overline P_1^{\rm exp}$ =& $(0.441 \pm 0.004 \pm 0.013)~{\rm GeV}^{-2}$~, \\
$\overline P_2^{\rm exp}$ =& $(-0.440 \pm 0.005 \pm 0.013)~{\rm GeV}^{-2}$~,
\end{tabular} 
\eeqa
which are in good agreement with the LET predictions. From our
fourth order results, we get for the sum of renormalized Born, third 
and fourth order loop and counterterm contributions
\beqa
\overline P_1 &=&\,\,\,\, (0.460 + 0.017 - 0.133 + 0.0048\,\xi_1)~{\rm 
GeV}^{-2}~, \\
\overline P_2 &=& -(0.449 + 0.058 - 0.109 + 0.0048\,\xi_2)~{\rm GeV}^{-2}~,
\eeqa
where the $\xi_{1,2}$ only depend on the $N\Delta$ transition magnetic moment.
We note the rather sizeable (25\%) correction from the fourth order
loops which at first sight seems to destroy the agreement between the LETs
and the data. However, it is known that $\kappa^* \simeq 4\ldots 6$, so we collect
in table~\ref{tab:kap} the predictions for $\overline{P}_{1,2}$ for reasonable
variations of $\kappa^*$. We see that for $\kappa^* = 4$, the delta contribution
almost completely cancels the large fourth order loop effect and thus the
predictions for the P--wave slopes are within 7\% of the empirical
values. Note, however, that the empirical finding $|\overline P_1^{\rm
  exp}| = |\overline P_2^{\rm exp}|$ is difficult to reconcile with any theory.

\renewcommand{\arraystretch}{1.2}
\begin{table}[hbt]
\begin{center}
\begin{tabular}{|l|c|c|}
    \hline
     $\kappa^*$  & $\overline P_1$ [GeV$^{-2}$] & $\overline P_2$ [GeV$^{-2}$]  \\ 
    \hline
  4.0                       &  0.408           &  $-$0.475       \\
  4.5                       &  0.416           &  $-$0.487       \\
  5.0                       &  0.427           &  $-$0.498       \\
  5.5                       &  0.439           &  $-$0.509       \\
  6.0                       &  0.450           &  $-$0.521       \\
  6.5                       &  0.461           &  $-$0.532       \\
\hline\hline
\end{tabular}
\centerline{\parbox{9cm}{
\caption{Prediction for the P--wave slopes for variations of the 
          $N\Delta$ transition magnetic moment $\kappa^*$.\label{tab:kap}}}}
\end{center} 
\end{table}   

\section{Summary}
\label{sec:sum}

In this paper, we have studied near threshold neutral pion photoproduction 
off protons in the framework of heavy baryon chiral perturbation theory
to complete one loop (fourth order) accuracy,
updating and extending previous work on this topic~\cite{bkmzpc,bkmlet}.
The pertinent results of this investigation can be summarized as follows:

\begin{enumerate}

\item[(i)] We have given the fourth order corrections (loops and counterterms)
to the three P-wave multipoles $P_{1,2,3}$. Two new low--energy constants
appear,  one for $P_1$ and the other for $P_2$.  We have also given 
analytic expressions for the corrections to the low--energy theorems 
for the P--wave slopes $\overline{P}_{1,2}$, see eqs.(\ref{LETloop41},\ref{LETloop42}).

\item[(ii)] We have analyzed the new MAMI data~\cite{schmidt} first in the
same approximation as it was done in previous works (i.e. the P--waves
to third order only). Using a realistic two parameter model for the
energy dependence of the electric dipole amplitude $E_{0+}$, we have
extracted the strength of the unitary cusp which agrees with the prediction
based on the final state theorem. 

\item[(iii)] Using the full one loop amplitude, one can fit the cross section
data and the photon asymmetry. The combination of S--wave LECs is stable
and agrees with previous determinations, leading to  $E_{0+} (\omega_{\rm thr})
= -1.1 \cdot 10^{-3}/M_{\pi^+}$. Two of the three P--wave LECs are
not well determined because of strong correlations. More photon asymmetry data
are needed to cure this situation.

\item[(iv)] We have analyzed the new LECs in the framework of resonance saturation
in terms of (dominant) $\Delta$--isobar and (small) vector meson excitations.
The isobar contributions depend only on the strength of the $N\Delta$ transition
magnetic moment. We have shown that for reasonable values of this constant,
the 25\% fourth order loop corrections to the P--wave LETs 
are almost completely cancelled by the
isobar terms. This solidifies the third order LET predictions, which are in 
good agreement with the data, cf. eqs.(\ref{LETt},\ref{LETe}).

\end{enumerate}

\vfill


\section*{Acknowledgements}

We are grateful to Reinhard Beck and Axel Schmidt for communicating the
new MAMI data before publication.

\newpage

\end{document}